\documentstyle[psfig,conf_iap]{article}
\begin{document}
\heading{%
%
The Redshift Distribution of Extragalactic Radio Sources}
%
\par\medskip\noindent
\author{%
C A Jackson$^1$, J V Wall$^2$
}
\address{%
Astrophysics, School of Physics, University of Sydney, NSW 2006, Australia
}
\address{%
Royal Greenwich Observatory, Madingley Road, Cambridge, CB3 OEZ, UK
}

\begin{abstract}

Extragalactic radio sources are a unique cosmological probe 
in that they trace large-scale structure on scales inaccessible 
to other wavelengths. However as radio survey data is inherently 2$D$, 
the redshift distribution, $N(z)$, is necessary 
to derive spatial information. To obtain this distribution either we
measure thousands of radio source redshifts to directly determine
$N(z)$ {\it or} we derive $N(z)$ from statistical analyses of 
radio source count and identification data.

In this paper we show how the dual-population unification scheme
can be incorporated into a rigorous statistical analysis of radio
source count data, with the result that our simple parametric evolution and
beaming model revises previous estimations of $N(z)$, 
specifically at low flux densites.
This revision is particularly pertinent given that
the new generation of radio surveys
extend to milli-jansky flux density levels: sampling source
densities high enough to reveal spatial structure.
In turn, these new radio surveys will provide potent tests which
will refine our model.

\end{abstract}
\section{Extragalactic radio sources as cosmological probes}

The development of large-scale structure had to take place at epochs
coresponding to redshifts greater than 0.2; moreover structure is now
known to extend beyond scales of $\sim200 h^{-1}$ Mpc. Both factors place
the extent and evolution of large-scale structure beyond the reach of
current optical and IR surveys and we must look outside these bands to
address such key cosmological issues. 

It has now been established that {\it extragalactic radio sources} can
trace structure both to very early epochs and on very large scales 
\cite{CR},\cite{LO},\cite{MM}. This is a direct
result of their uniform selection function: away from the Galactic plane
extragalactic radio sources are visible to very high redshifts ($z > 4$) 
with detection unaffected by obscuration. Moreover the high radio
luminosities and strong space-density evolution yield a peak selection
range of very high redshifts ($1 < z < 3$). 

It was \cite{LO} who
showed how to determine the 3D 2-point spatial
correlation function from 2D information in radio surveys themselves, a
cosmological Limber's equation, and an estimated redshift distribution
for the radio sources to the survey limit. This and more recent analyses
use the best estimates of $N(z)$ available, namely those from the
comprehensive analysis of \cite{DP}, in which all the
then-known redshift and source-count data were synthesized into a
determination of the epoch-dependent luminosity functions. These in turn
can be used to predict $N(z)$ at any frequency and flux-density level. The
statistical accuracy becomes rapidly poorer below flux densities
equivalent to $S_{1.4 {\rm \thinspace GHz}}$ = 100 mJy, 
as the results are then an
extrapolation.  Moreover the analysis considers steep-and flat-spectrum
populations separately and independently, while not including at all the
starburst galaxy population which dominates at levels below a few mJy.

Our recent analysis of radio source evolution in terms of a
dual-population unified model \cite{WJ},\cite{JW}
describes how the flat and steep-spectrum populations are
physically related, and includes this latter starburst population.
Accordingly it promises to predict $N(z)$ with somewhat greater
reliability, particularly at the lower flux densities. 


\section{Evolution and beaming of the radio source populations}

Our dual-population unified model is based on the two Fanaroff-Riley
(1974) classes of radio galaxies as parent populations. Anisotropic
radiation mechanisms of relativistic beaming along the radio axes and
dusty tori shrouding the nuclei result in core-dominated quasars and
BL\,Lac objects from these, when radio axes coincide closely with our
line-of-sight. We determined the cosmic evolution history of the two
populations FRI and FRII from low-frequency (151 MHz) survey statistics. 
We then determined beaming models for the related quasars and BL\,Lac
objects by matching predicted and observed source counts at 5 GHz using
Monte Carlo orientation of the parent population.  In this process we
adopted the pure luminosity evolution model for the starburst galaxy
population derived by \cite{SA}.  In the process we found that
a combination of strong cosmic evolution of the FRII sources coupled with
no evolution of the FRI population is required to fit the low-frequency
count data, while the best-fit beaming models for the two populations
have parameters, jet-speeds in particular, which match those observed in
VLBI observations of individual sources. The composite model which we
found for evolution and beaming gave a natural explanation of the change
in source count shape with frequency, while showing good agreement with
several other independent data sets.

\section{The redshift distribution of radio sources {\boldmath $N(z)$}}

Our evolution and beaming model can be used predict at any frequency the
intensity-dependent mix of the three underlying radio-source populations:
the starburst galaxies, and the FRI and FRII radio galaxies together with
their beamed (on-axis) counterparts. Figure 1 shows this mix for 1.4 GHz.

The model can also be used to predict the redshift distribution $N(z)$
for any frequency and in a given intensity range. Figure~2 shows the
$N(z)$ derived for 1.4 GHz in the flux-density range $1 < S_{1.4 GHz} <
100$ mJy. The dominant populations in this range (see Figure 1) are the
FRI sources (beamed and unbeamed versions) at moderate redshifts, with
starburst galaxies at low redshift; the high-power FRII sources make only
a minor contribution. The implication for the latest generation of radio
surveys which extend to $S_{1.4 GHz} \sim 1$~mJy is that these surveys
{\it completely} sample this the most powerful radio-source population;
one entire population is uniformly sampled across its entire evolutionary
history to the highest observable redshifts by these sensitive radio
surveys.

The difference between this estimation of $N(z)$ and those used
previously - in particular those from the models of \cite{DP}
 - can be attributed to (i) the `spike' at $z \sim 1$ due to the
starburst galaxies, a population not included in previous analyses, and
(ii)  the lower median redshift due to the dominance of the unevolving
FRI population at this level.

\vspace*{0.2in}

\begin{center}
{\bf Table 1.} Deep radio surveys \\
\vspace*{0.05in}
\begin{tabular}{lccc}
\hline
\\
 & FIRST \cite{BB} & NVSS \cite{CC} & SUMSS \cite{HH} \\
\\
\hline
\\
Frequency & 1400 MHz & 1400 MHz & 843 MHz \\
Area (sq deg) & 10,000   & 33,700 & 8,000 \\
Resolution ($''$) & 5$''$ & 45$''$ & 43$''$  \\
Detection limit  & 1 mJy & 2.5 mJy & $\sim$5 mJy \\
Coverage & NGP &   $\delta >$ $-$40$^{\circ}$ &
 $\delta <$ $-$30$^{\circ}$ \\
Sources / sq deg & $\sim$90 & $\sim$60 & $\sim$40 \\ 
\hline
\end{tabular}
\end{center}

%
\begin{center}
\begin{minipage}{8cm}
\hspace*{-0.4in}
\vspace*{-0.2in}
\psfig{figure=jacksonF1.ps,height=8cm,bbllx=20pt,bblly=20pt,bburx=400pt,bbury=600pt}
\end{minipage}
\vspace*{0.15in}
\end{center}
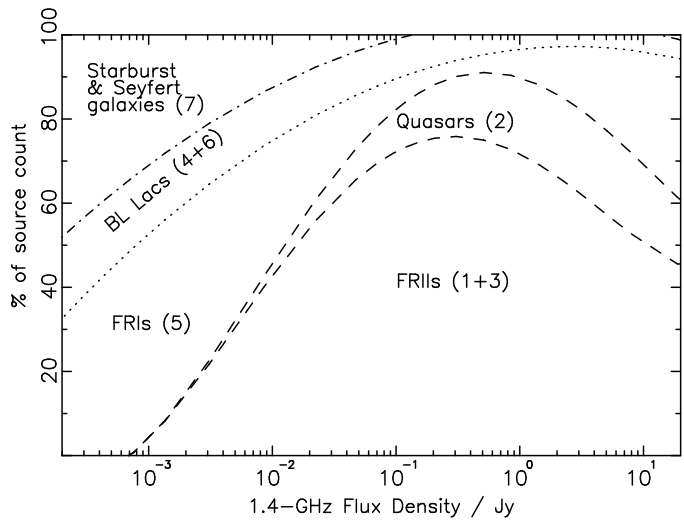
{\bf Figure 1.}{\small The predicted integral population mix at 1.4 GHz from
our dual-population unified model \cite{WJ}, \cite{JW}.  The
contribution from the FRII population (FRII radio
galaxies and quasars) is almost negligible at
$S_{1.4 {\rm \thinspace GHz}}$=1 mJy.} 

\begin{center}
\begin{minipage}{7cm}
\hspace*{-0.4in}
\psfig{figure=jacksonF2.ps,height=8cm,bbllx=20pt,bblly=20pt,bburx=400pt,bbury=600pt}
\end{minipage}
\vspace*{0.15in}
\end{center}
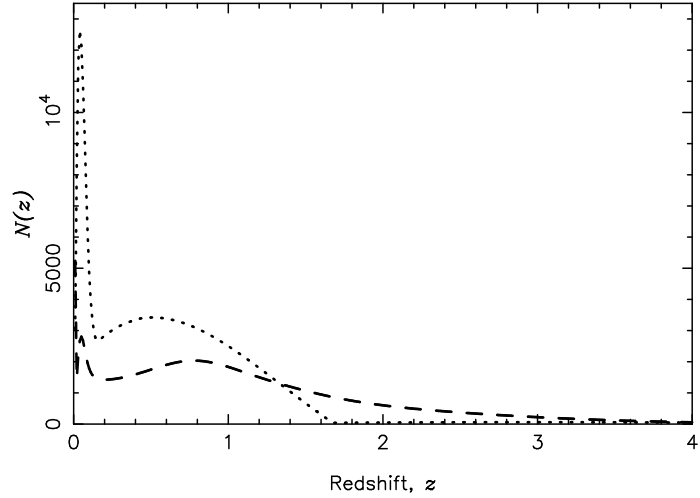
{\bf Figure 2.}{\small The total predicted $N(z)$  
for sources in the flux density range 1 $\le S_{1.4 {\rm \thinspace
GHz}} \le$ 100 mJy (solid line) from evolution models of 
\cite{DP} (average of models 1-4,6 \& 7) (dashed line), 
and \cite{JW} (dotted line).}

\vspace*{0.15in}

Our evolution model (and in consequence the $N(z)$ predictions) will be
further refined by (i) incorporating the results from a multi-object
spectroscopy campaign (WYFFOS + WHT, 2dF + AAT) which targets FIRST
survey sources, and (ii) redefining the local radio luminosity function
using spectra obtained by the 2dF galaxy redshift survey (M Colless,
these proceedings). 

\vspace*{-0.1in}


%
\vfill

\begin{iapbib}{99}{

\bibitem{CR} Cress C M, Helfand D J, Becker R H, Gregg M D, White R L,
  1996, ApJ, {\bf 473},~7

\bibitem{LO} Loan A J, Wall, J V, Lahav, O, 1997,
MNRAS, {\bf 286}, 994

\bibitem{MM} Magliocchetti M, Maddox S J, Lahav O, Wall J V, 1998,
  MNRAS, {\it in press}

\bibitem{DP} Dunlop J S, Peacock J A, 1990, MNRAS, {\bf 247}, 19

\bibitem{WJ} Wall J V, Jackson C A, 1997, MNRAS, {\bf 290}, L17

\bibitem{JW} Jackson C A, Wall J V, 1998, MNRAS, {\it in press}

\bibitem{SA} Saunders W, Rowan-Robinson M, Lawrence A, Efstathiou G,
Kaiser N, Ellis R S, Frenk C S, 1990, MNRAS, {\bf 242}, 318

\bibitem{BB} Becker R H, White R L, Helfand D J, 1995, ApJ, {\bf 450}, 559

\bibitem{CC} Condon J J, Cotton W D, Greisen E W, Yin Q F, Perley R A,
Taylor G B, Broderick J J, 1998, A J, {\bf 115}, 1693

\bibitem{HH} Hunstead R W, 1998, {\it Observational 
Cosmology from the New Radio Surveys},
{\it eds}, Bremer M N, Jackson N J \& Perez-Fournon I, Kluwer
Academic Publishers

%
}
\end{iapbib}
\end{document}